# A Metamodel of the Telemac Errors

Fabrice Zaoui[1], Cédric Gœury[1], Yoann Audouin[1]
[1] EDF R&D – National Laboratory for Hydraulics and Environment (LNHE)
6 quai Wattier, 78401 Chatou, France
E-mail : fabrice.zaoui@edf.fr

*Abstract—* **A Telemac study is a computationally intensive application for the real cases and in the context of quantifying or optimizing uncertainties, the running times can be too long. This paper is an example of an approximation of the Telemac results by a more abstract but significantly faster model. It shows how a metamodel can be easily built with low computational costs, and how it can help to understand and improve some global results of Telemac.**

## I. Introduction

Many sources of uncertainty lie in the real-world problems. Telemac as any model (i.e. approximation of reality) is error prone since uncertainties appear in the initial or boundary conditions, the system parameters, the modelling simplification or the numerical calculations themselves. Therefore, it is difficult to say with confidence if the design of a Telemac model has met all the requirements to be optimal. Calibration consists of tuning the model parameters so that the results are in better agreement with a set of observations. This phase is crucial before any further study can be conducted by avoiding a meaningless analysis or prediction based on false or too inaccurate results. This paper presents a statistical calibration of a Telemac 2D model (Gironde Estuary in France) with the learning of Telemac errors by a metamodel (i.e. a model of the simulation errors) to make the best use of limited observations data over a short time period. The metamodel here is a simplified version of Telemac behaving the same for all the locations where observation points are available. If the metamodel is correct, it will be able to compute as Telemac would do but with a highly reduced computational cost.

While doing an analysis of the simulation errors it is shown how a metamodel (or surrogate, i.e. methods like Kriging, Polynomial Chaos, Neural Networks, etc.) can be designed with a low number of calls to Telemac. The minimal number of calculations to have a good approximation can be empirically defined [1][2][3] by:

$$N_{calls} \cong 10 \times N_{parameters}$$

The strength of the metamodel is highlighted with the reduction of the modelling discrepancy between the simulation and the observations by considering different errors globally or locally depending on the goal(s) to achieve, see Fig. 1. For instance, a multi-objective calibration can be quickly conducted with a metamodel even when the optimizer is a meta-heuristic requiring many runs like a NSGA-II based algorithm [4]. The optimal solution obtained after many calls to the metamodel is systematically validated by checking its expected performance with a last call to Telemac.

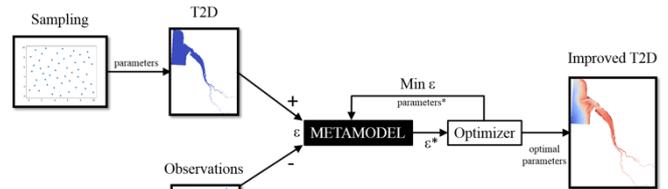

Figure 1. Workflow of the metamodel use

## II. Forward Model

Telemac 2D is used to solve the shallow waters equations on the real case of the Gironde Estuary in southwestern France, see Fig. 2. This estuary is important because of its large size involving many economic and environmental considerations.

Several Telemac studies of the estuary have been carried out for the last decade mainly concerning the hydrodynamic [5][6] and morphodynamic [7][8][9]. In these studies, the importance of the calibration phase is systematically emphasized to achieve operational performance. Are concerned the Telemac computation time and robustness but also the accuracy of the water levels for tide forecasts.

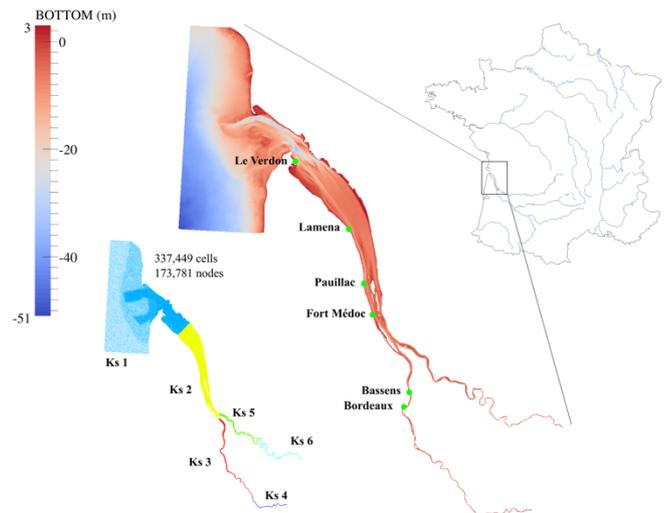

Figure 2. Gironde estuary with Telemac. Location of observation stations, bathymetry, friction zones and mesh size

The Gironde is a navigable estuary formed from the junction of rivers Dordogne and Garonne just downstream of the center of Bordeaux city, it is the largest estuary in western Europe. The hydraulic model used in this work covers approximatively 195 km between the fluvial upstream and the marine boundary conditions downstream representing an area of around 635 km². The mesh of 173,781 nodes varies with cell lengths of 40 m within the areas of interest (navigation channel) to 750 m offshore (western and northern sectors of the model). This model has been built by CEREMA in the framework of the Gironde XL project [10].

The boundary condition along with the marine border of the model has been set up using depth-averaged velocities and water levels coming from the dataset of Legos numerical model TUGO (46 harmonic constants). Surge data, describing the difference between the tidal signal and the observed water level, is taken into account using results from the Hycom2D model of SHOM [11]. Surface wind data is also considered for simulating flows under windy conditions.

The period of interest concerns two days of August 2015 with six stations providing water levels every minute [6]. The time step for Telemac is 10 seconds and the ratio between the simulated and execution times is about 120 on 56 Intel Xeon cores @ 2.4 GHz.

The calibration will aim to decrease water level differences between Telemac and the observation stations by optimizing some Root Mean Squared Errors over the two-day period:

$$J_s(x) = \sqrt{\frac{1}{N_t} \sum_{i=1}^{N_t} (T_i^s(x) - O_i^s)^2} \quad (1)$$
$$\forall s \text{ in } (1..N_s)$$

where:
- $N_s$ is the number of stations;
- $N_t$ is the number of time steps;
- $J_s$ is the RMSE for station $s$;
- $x$ is a vector of physical parameters;
- $T$ is the water level computed by Telemac;
- $O$ is the corresponding observation.

If $J_s(x)$ is minimized for a certain value of $x = x_{opt}$, the Telemac model is assumed to be well calibrated for the station $s$ on average over two days. Finding $x_{opt}$ implies the use of a specific algorithm with many trials and calls to Telemac to define an optimality path. These repeated model evaluations put a heavy workload on CPUs, and it makes more sense to replace $J_s(x)$ by an estimate $\hat{J}_s(x)$ with the help of a metamodel to be less resource intensive, see Fig. 1.

Therefore, the problem can be written as a bound constrained optimization:

$$\begin{aligned} minimize \quad & \hat{J}_s(x) \cong J_s(x) \\ subject\ to \quad & x_{lb} \leq x \leq x_{ub} \end{aligned} \quad (2)$$

where the vector $x$ is made of three tidal parameters $(\alpha, \beta, \gamma)$ and six friction coefficients $K_{S_i}$ for $i$ in $(1..6)$, see Fig. 2, with bound constraints on the values of variables.

## A. Tidal parameters

Tidal characteristics are imposed using a database of harmonic constituents to force the open boundary conditions. For each harmonic constituent, the water depth $h$ and horizontal components of velocity $u$ and $v$ are calculated, at point $M$ and time $t$ by:

$$\begin{cases} F(M,t) = \sum_i F_i(M,t) \\ F_i(M,t) = f_i(t) A_{F_i}(M) \cos\left(\frac{2\pi t}{T_i} - \phi_{F_i}(M) + \varphi_i^0 + w_i(t)\right) \end{cases} \quad (3)$$

where $F$ is either the water level (referenced to mean sea level) $z_s$ or one of the horizontal components of velocity $u$ or $v$, $i$ refers to the considered constituent, $T_i$ is the period of the constituent, $A_{F_i}$ is the amplitude of the water level or one of the horizontal components of velocity, $\phi_{F_i}$ is the phase, $f_i(t)$ and $w_i(t)$ are nodal factors and $\varphi_i^0$ is the phase at the original time of the simulation.

The water level and velocities of each constituent are then summed to obtain the water depths and velocities for the open boundary conditions:

$$\begin{cases} h = \alpha \sum_i z_{s_i} - z_f + z_{mean} + \gamma \\ u = \beta \sum_i u_i \\ v = \beta \sum_i v_i \end{cases} \quad (4)$$

where $z_f$ is the bottom elevation and $z_{mean}$ the mean reference level. In (4), the multiplier coefficients of the tidal range and velocity, respectively $\alpha$ and $\beta$, at boundary locations and the sea level correction $\gamma$ are the coordinates of $x$ in (1) corresponding to tidal parameters.

Upper and lower bounds $(x_{lb}, x_{ub})$ for the tidal parameters are [6]:

TABLE 1: BOUNDS FOR TIDAL PARAMETERS

| Parameter | Lower bound $x_{lb}$ | Upper bound $x_{ub}$ |
|---|---|---|
| $\alpha$ | 0.8 | 1.2 |
| $\beta$ | 0.8 | 1.2 |
| $\gamma$ | 0.35 | 0.54 |

## B. Friction coefficients

Friction terms come into the momentum equation of the shallow system are treated in a semi-implicit form in Telemac 2D. The bed friction in $X$ and $Y$ directions can be written as a function of the Strickler coefficient $K_s$:

$$\begin{aligned} F_X &= \frac{u\sqrt{u^2+v^2}}{K_s^2 h^{4/3}} \\ F_Y &= \frac{v\sqrt{u^2+v^2}}{K_s^2 h^{4/3}} \end{aligned} \quad (5)$$

where $h$ is the water depth and $(u, v)$ the velocity components. Bounds for the friction coefficients are [6]:

TABLE 2: BOUNDS FOR FRICTION COEFFICIENTS

| Parameter | Lower bound $x_{lb}$ | Upper bound $x_{ub}$ |
|---|---|---|
| $K_{S1}$ | 30 | 46 |
| $K_{S2}$ | 64 | 96 |
| $K_{S3}$ | 80 | 100 |
| $K_{S4}$ | 20 | 30 |
| $K_{S5}$ | 64 | 96 |
| $K_{S6}$ | 36 | 54 |

### III. METAMODELLING

This part explains how to pass from $J_s(x)$ to $\hat{J}_s(x)$ in (2) by a training of the Telemac errors on the water levels given the space defined in Table 1 and 2.

*A. Definition*

A metamodel is a CPU time inexpensive mathematical function that can replace a more complex model by approximation. Indeed, many simulation codes have become accurate for complex problems in physics as with 3D modelling. On the other hand, the simulation times have not necessarily decreased, and this makes some engineering studies difficult if they require for example an uncertainty quantification, a sensitivity analysis or optimization that rely on these models. A way out of this difficulty is to build a metamodel [12] with a limited number of simulations, i.e. a low number of outputs from the original model.

Metamodels can be based on different methods like support vector machines, regression trees, proper orthogonal decomposition, neural networks, polynomial chaos, kriging and so on. Recent examples of metamodels with the Telemac-Mascaret system use the polynomial chaos [13][14][15] or Gaussian Process [15][16] for a steady state hydraulics.

*B. Design*

A design of experiments is the first step to build a metamodel. It is used to sample the space of input parameters. In this way, the space filling design has been developed to address the issues of numerical simulation. This approach tries to best distribute a minimal number of samples in the input parameter space in order to capture the non-linear behaviour of output parameters.

One of the most used design of experiments is the method called Latin Hypercube Sampling (LHS). This method generates well-distributed and non-redundant samples. It also addresses a wide variety of problems in terms of size of the input and output parameter spaces [17].

Fig. 3 shows two designs of experiments for a simple case with 50 points. The first design on the left is based on a random sampling while the second on the right is based on LHS. Input parameters space is better explored with LHS. For the case of the Gironde estuary, the LHS is built with 90 points as the length of $x$ is 9. This is a minimal size according to [1][2][3].

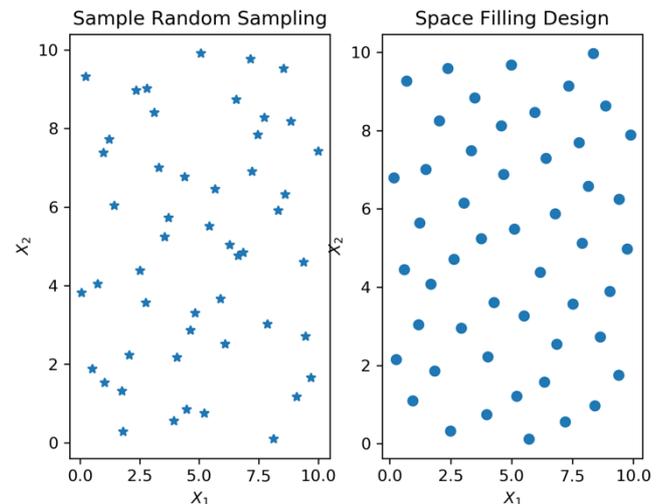

Figure 3. Comparison of two designs of experiments

*C. Kriging Example*

Kriging is method of geostatistics, a family of techniques building response surfaces from a limited number of samples and estimating values at unknown locations. This method is named in honor of engineer D. G. Krige [18] by G. Matheron [19] who formalized the approach a few years later.

The method approximates the deterministic function $y$ by $\hat{y}$ as the sum of a trend and a stationary stochastic process:

$$\hat{y} = \sum_{i=1}^{k} \lambda_i g_i(x) + Z(x) \quad (6)$$

$$cov[Z(x_i), Z(x_j)] = \sigma^2 R(x_i, x_j)$$

where:
- the trend $\sum_{i=1}^{k} \lambda_i g_i(x)$ can be seen as a regression model with $\lambda$ the regression parameter vector and $g_i$ the basis functions;
- $Z(x)$ is a realization of a stochastic process with zero mean and spatial covariance function;
- $\sigma^2$ is the process variance;
- $R$ is a correlation function;
- $k$ is the number of samples from LHS ($k = 90$).

*D. Validation*

An out-of-sample testing is used to validate the Kriging for assessing how the metamodel performs for unknown data. Ten points are added for testing the metamodel. Looking at the mean error over the six stations, the metamodel interpolates the test data with a mean squared error of $2.22 \cdot 10^{-5}$ m$^2$ and a coefficient of determination $R^2$ of 0.996. Thus, the quality of the metamodel as estimator is considered satisfactory.

### IV. APPLICATIONS

Paragraph § *A* is a distribution analysis of the errors in (1) after propagating the LHS design with Telemac. The paragraphs § *B* and § *C* are two applications of the metamodel that require many function evaluations $\hat{J}_s(x)$.

It should be noted here that the time required for the construction or use of Kriging is not significant compared to the calculation time of a Telemac run. Thus, the repeated metamodel evaluations are not an issue.

*A. Error Analysis*

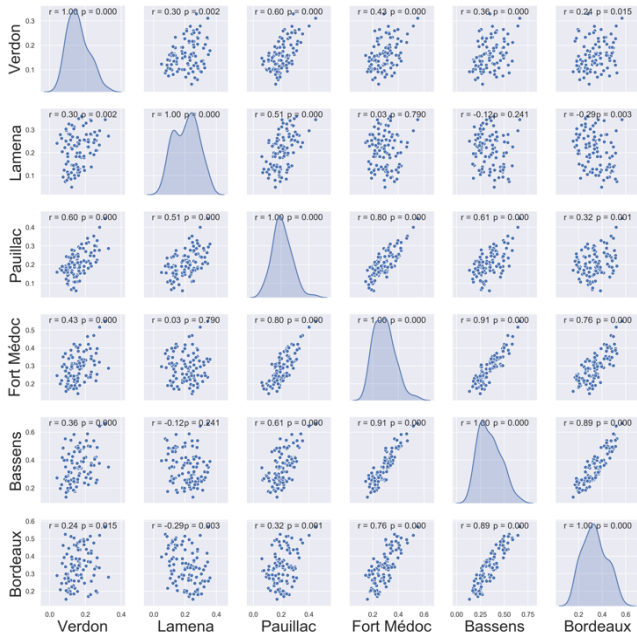

Figure 4. Scatter plot matrix of $J(x)$ for the LHS

Fig. 4 plots the distribution of RMSE in (1) after propagating the LHS. For each station, the distribution looks like a Gaussian and two parts can be clearly identified depending on the geographical location and the distance from the sea, see Fig. 2. The errors on the Fort Médoc, Bassens and Bordeaux stations have a similar behaviour and clearly different from other stations closer to the sea.

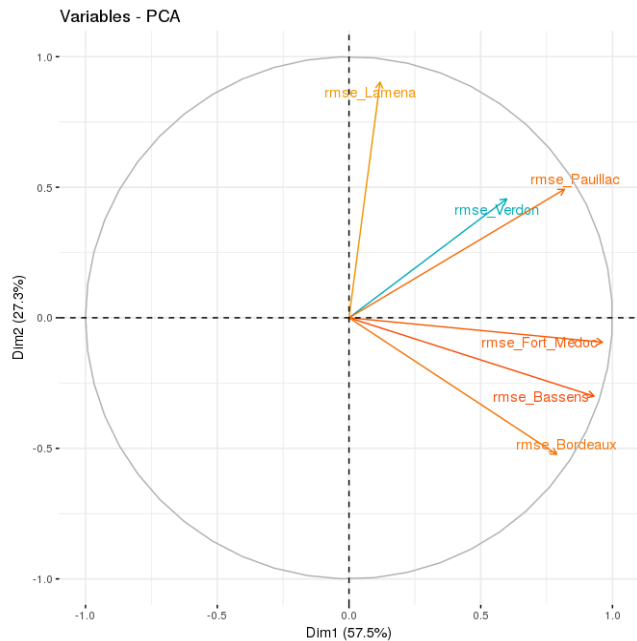

Figure 5. PCA correlation circle of $J(x)$ for the LHS

A Principal Component Analysis (PCA) is a method of data mining that consists of reducing the dimensionality of the problem to extract information. It is a projection method into a smaller space that decreases the number of variables. A graphical diagnostic is possible on the reduced dimensions. Fig. 5 is a diagnostic example with the correlation circle of PCA for the two first dimensions (i.e. the most significant ones, 85% of the variance is explained here). Contributions to the dimensions differ according to the position of the stations. The group consisting of Fort Médoc, Bassens and Bordeaux is confirmed. The errors in this group do not have the same behaviour as the other stations. In particular, the behaviour of the Lamena station errors is orthogonal to that of Fort Médoc, which makes the global approaches less relevant.

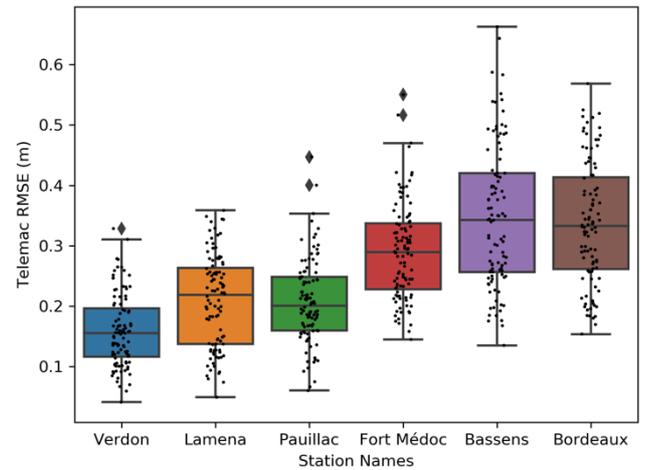

Figure 6. Box plot of $J(x)$ for the LHS

Fig. 6 shows the distribution of RMSE in (1) with quantiles. Errors are larger on average for the most inland stations. It should be noted that in this study only tidal and friction parameters are considered. The influence of the hydrological forcing of the two upstream rivers Garonne and Dordogne is not studied. This one can have an influence at least locally [5].

*B. Sensitivity Analysis*

A global sensitivity analysis is carried out with a variance decomposition using the Sobol' indices [20]. The analysis makes it possible to rank in order of importance the sources of uncertainty on the output $\hat{J}_s(x)$ of the metamodel.

The first order Sobol' index of variable $x_i$ (i.e. linear effect of $x_i$ on $\hat{J}_s$) is:

$$S_i = \frac{\mathbb{V}_i}{Var(\hat{J}_s)} \quad (7)$$
$$\text{with } \mathbb{V}_i = Var[\mathbb{E}(\hat{J}_s|x_i)]$$

The second order Sobol' index of two variables $x_i$ and $x_j$ (i.e. cross influence on $\hat{J}_s$) is:

$$S_{ij} = \frac{\mathbb{V}_{ij}}{Var(\hat{J}_s)} \quad (8)$$
$$\text{with } \mathbb{V}_{ij} = Var[\mathbb{E}(\hat{J}_s|x_ix_j)] - \mathbb{V}_i - \mathbb{V}_j$$

And the total Sobol' index of variable $x_i$ (linear and non-linear effects on $\hat{J}_s$) is:

$$S_{Ti} = S_i + \sum_j S_{ij} + \sum_{j<k} S_{ijk} + \cdots = 1 - S_{\sim i} \quad (9)$$

The computation of Sobol' indices is often done with a numerical integration based on Monte Carlo method. This method is inexpensive here because the metamodel is extremely fast.

Sobol' indices are computed on the variance of the mean value RMSE over all the stations $\hat{J}(x)$:

$$\hat{J}(x) = \frac{1}{N_s} \sum_{s=1}^{N_s} \hat{J}_s(x) \quad (10)$$

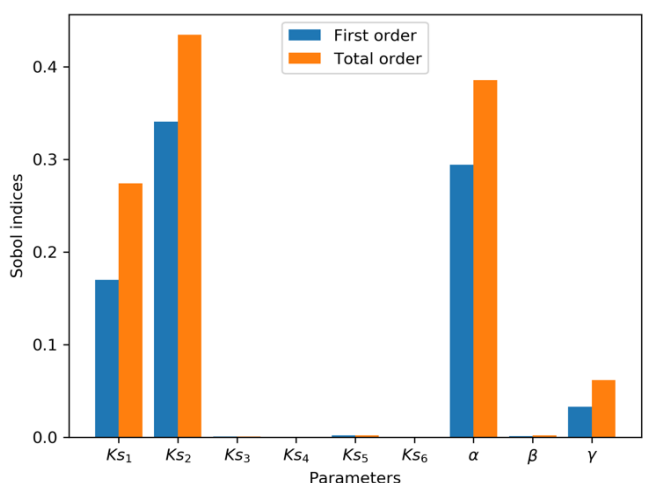

Figure 7. Sobol' indices (total and first order) of $x$ on $\hat{J}(x)$

According to Figure 7, only 4 of the 9 parameters are significant. The other parameters could have been ignored (i.e. fixed) because on average they have no influence on the water level errors.

This result is in perfect agreement with the transient computation of Sobol' indices as mentioned in [6].

*C. Optimization*

Sensitivity analysis, calibration, design, reliability assessment and other studies require many function evaluations. In this section, several calibration examples using metamodel and optimization are presented.

The first example is the minimization of $\hat{J}(x)$ in (10). This is done with two different optimization algorithms, a metaheuristic (PSO [6]) and a gradient based method (Newop, available in the sources of the Telemac-Mascaret system). Newop requires less function evaluations than PSO to converge but its final result can depend on the initial guess.

Fig. 8 shows the convergence of PSO and Newop with the values of $\hat{J}(x)$ minimized. Both algorithms give exactly the same optimal result after convergence: $0.1259\ m$. This result is an average value of all the stations.

To see if the metamodel is reliable enough, the corresponding optimal solution for $x$ has been checked directly with Telemac. That one gives: $0.1265\ m$. The difference is 0.5%, less than $1\ mm$ which is remarkable and confirms the good construction of the metamodel.

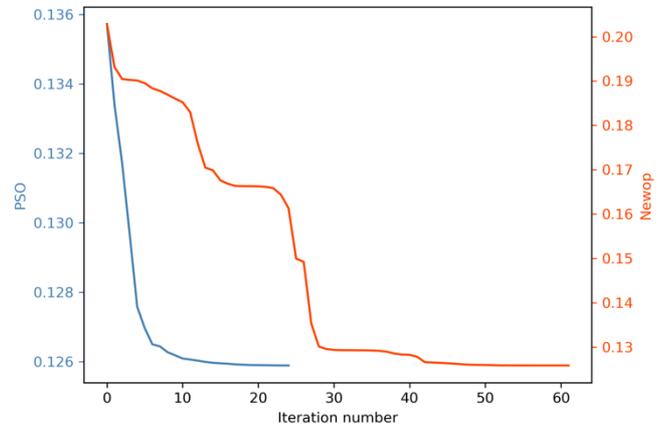

Figure 8. Convergence of the optimization algorithms

The result of this optimization for one period of a tidal wave is illustrated in Fig. 9. It corresponds to the worst result at Bordeaux station with an error value greater than $17.5\ cm$ on two days.

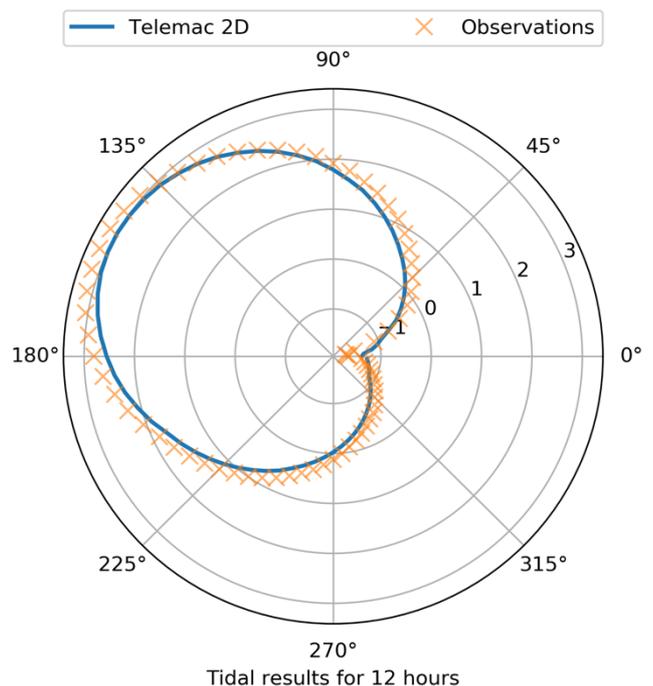

Figure 9. Tidal wave at Bordeaux after calibration (average goal)

Instead of looking for an average goal for all the stations, it might be interesting to calibrate each station individually, that is, to minimize $\hat{J}_s(x)$ in (2) instead of $\hat{J}(x)$ in (10).

In Fig. 10 improvements per station are indicated in comparison of the previous optimization. For each station there is a different optimal solution for $x$ (tidal and friction parameters) that minimizes the RMSE on a two-day period.

As might be expected, the optimization of a particular station is locally more interesting than the shared solution by all the stations.

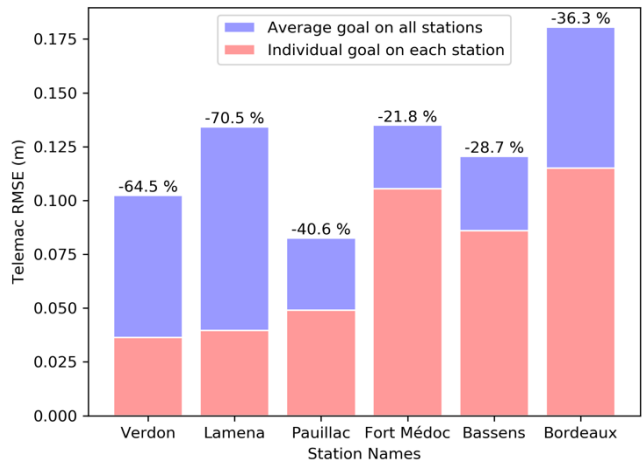

Figure 10. Improvement with an individual goal

One can also think of another mono-objective like minimizing the standard deviation if the model must have the same performance whatever the station or minimizing the maximum error which is a non-differentiable problem but treatable with PSO as shown in Fig. 11.

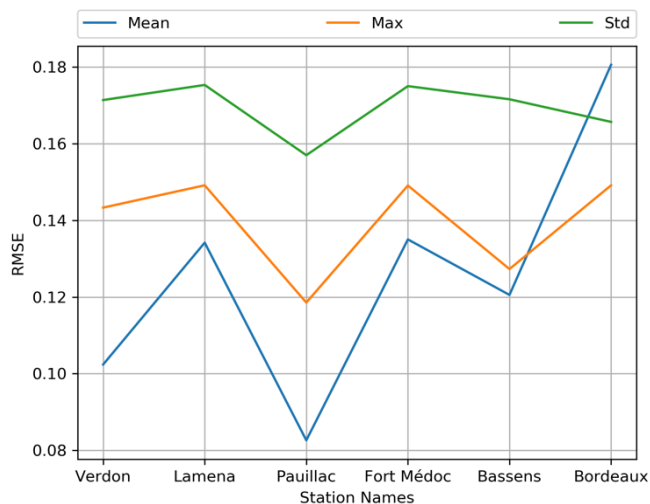

Figure 11. Minimizing $\hat{J}_s(x)$ for three different objectives

RMSE for Fort Médoc and Lamena stations are orthogonal, see Fig. 5. Tackling the minimization of these two errors in the same mono-objective problem is risky because minimizing one is in conflict with the other. It is better to deal with two separated objectives simultaneously; it is the role of the multi-objective optimization. If the problem is nontrivial, it will find a number of nondominated solutions defining a Pareto front. This frontier is a trade-off between objectives.

The multi-objective evolutionary algorithms are dedicated to find a set of Pareto optimal solutions. This family of algorithms requires a high number of function evaluations to converge but thanks to the metamodeling this drawback is no more an issue. One of the most used algorithms is NSGA-II [4]. The solutions of a Pareto front with NSGA-II for the simultaneous minimization of RMSE for Fort Médoc and Lamena stations is presented in Fig. 12.

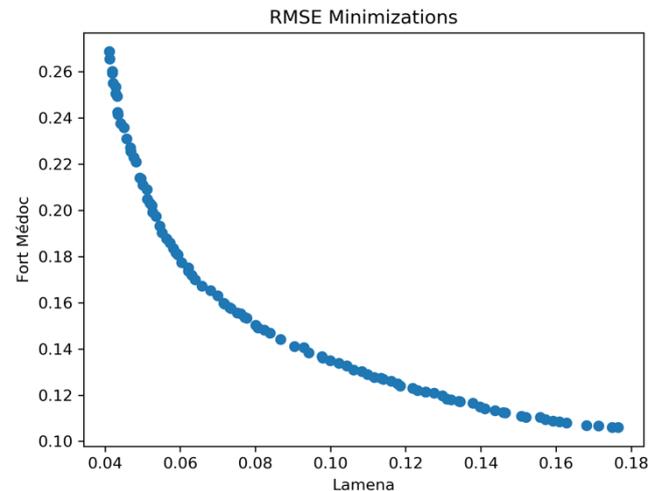

Figure 12. Pareto front of a two-objective minimization

With a multi-objective approach, it is easy to confirm that the objectives in Fig. 11 are conflicting. Indeed, if a two-objective minimization is launched with the pair (Max, Mean) or (Std, Mean), the result is again a Pareto front as illustrated in Fig. 13.

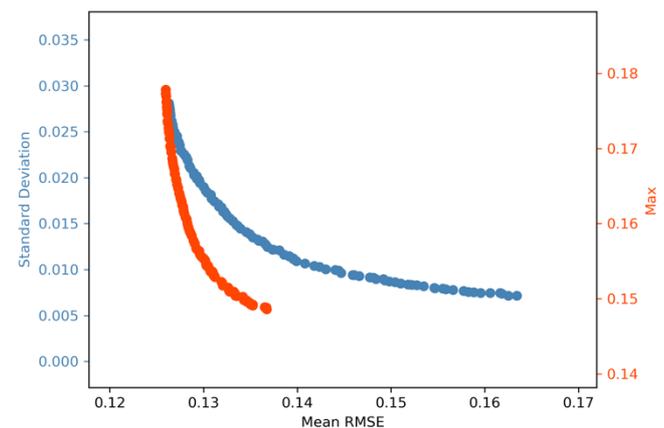

Figure 13. Pareto fronts of 2 two-objective minimizations

It is still possible to calibrate a model by supporting several objectives simultaneously, the number of objectives to be achieved may be greater than two. Other criteria than the RMSE could have been chosen, such as the bias or the Nash coefficient to evaluate the model efficiency. Following this idea, a new optimization with three objectives is carried out with always the metamodel based on the Kriging to evaluate the criteria. The results are shown in Fig. 14.

$$BIAS = \frac{1}{N_t}\sum_{i=1}^{N_t} T_i^s(x) - \frac{1}{N_t}\sum_{i=1}^{N_t} O_i^s \qquad (11)$$

$$NASH = 1 - \frac{\sum_{i=1}^{N_t}(T_i^s(x) - O_i^s)^2}{\sum_{i=1}^{N_t}(T_i^s(x) - \bar{O}^s)^2}$$

$$\forall s \; in \; (1..N_s)$$

From Fig. 14 it is clear that Nash values are not significative because they are all close to value 1. The bias is always positive with a value around $3\ cm$. Bias and RMSE may conflict with the lowest error values.

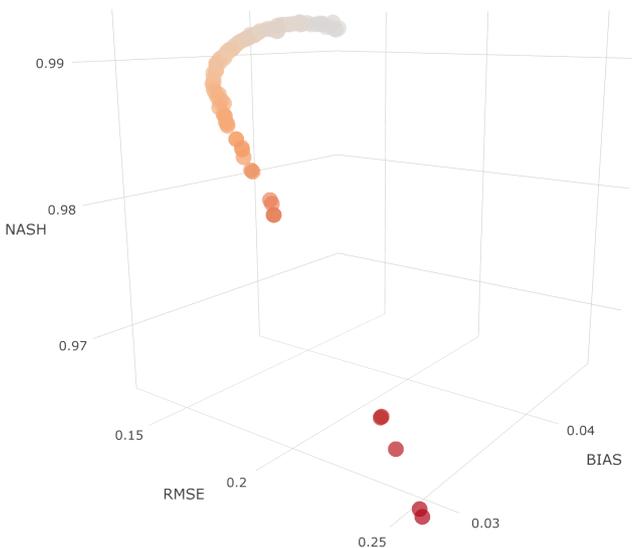

Figure 14. Pareto front of a three-objective minimization

## V. Conclusion

The purpose of this article is to introduce practitioners of the Telemac system to the construction of a metamodel. A metamodel is explained on the example of an approximation of Telemac errors compared to observations on a real case. The metamodel is then used for various applications concerning the analysis and the sensitivity of the errors as well as the calibration by minimization of these same errors.

A perspective of this work is to assess the validity period of a metamodel. Indeed, the behaviour of the physics can slightly change over time and it may be necessary to update the metamodel. This can be important for the implementation of data assimilation techniques.

## Acknowledgement

The authors gratefully acknowledge the open source community and especially that of the OpenTURNS (An Open source initiative for the Treatment of Uncertainties, Risks'N Statistics).

## References

[1] D. R. Jones, M. Schonlau and W. J. Welch, "Efficient Global Optimization of Expensive Black-Box Functions", Journal of Global Optimization, vol. 13, pp. 455-492, 1998

[2] A. Marrel, B. Iooss, B. Laurent, O. Roustant, "Calculations of Sobol indices for the Gaussian process metamodel", Reliability Engineering and System Safety, 94, pp. 742-751, Elsevier, 2009

[3] J. L. Loeppky, J. Sacks, W.J. Welch, "Choosing the sample sizeof a computer experiment: A practical guide", Technometrics, 51, pp. 366–376, 2009

[4] K. Deb, A. Pratap, S. Agarwal, T. Meyarivan, "A fast and elitist multiobjective genetic algorithm: NSGA-II", IEEE Transactions on Evolutionary Computation, vol. 6 (2), pp. 182-197, 2002

[5] V. Laborie, N. Goutal, S. Ricci, M. De Lozzo, P. Sergent, "Uncertainty Quantification for the Gironde Estuary Hydrodynamics with TELEMAC2D", Advances in Hydroinformatics, Springer, pp. 205-219, 2018

[6] C. Gœury, F. Zaoui, Y. Audouin, P. Prodanovic, J. Fontaine, P. Tassi, R. Ata, "Finding Good Solutions to Telemac Optimization Problems with a Metaheuristic", 25[th] Telemac-Mascaret User Conference, Norwich, England, pp. 159-167, October 2018

[7] C. Villaret, N. Huybrechts, L. A. Van, "Large Scale Morphodynamic Modeling of the Gironde Estuary", 18[th] Telemac-Mascaret User Conference, Chatou, France, pp. 117-123, October 2018

[8] C. Villaret, N. Huybrechts, A. G. Davies, "A Large Case Morphodynamic Process-based Model of the Gironde Estuary", Jubilee Conference Proceedings, NCK-Days, pp. 69-76, 2012

[9] N. Huybrechts, C. Villaret, F. Lyard, "Optimized Predictive Two-Dimensional Hydrodynamic Model of the Gironde Estuary in France", Journal of Waterway, Port, Coastal, and Ocean Engineering, vol. 138(4), pp. 312-322, 2012

[10] F. Klein, A. Fort, A. Sottolichio, A. Beudin, G. Mattarolo, C. Goeury, A. Ponçot, J.-P. Argaud, S. Orseau, P. Tassi, N. Huybrechts, S. Cai, H. Smaoui, R. Gasset, Y. Nedelec, V. Laborie, R. Leroux, S. Barthelemy, M. Ali, S. Kaidi, "Gironde XL : an example of how estuarine scientific research can improve dredging and navigation operations in ports", VII Congreso Nacional de la ATPYC, 4[th] Mediterranean Days, Sevilla, Spain, 2018

[11] E. P. Chassignet, H. E. Hurlburt, O. M. Smedstad, G. R. Halliwell, P. J. Hogan, A. J. Wallcraft, R. Baraille, R. Bleck, "The HYCOM (Hybrid Coordinate Ocean Model) Data Assimilative System", Journal of Marine Systems, vol. 65(1-4), pp. 60-83, 2007

[12] J. Kleijnen, R. Sargent, "A Methodology for Fitting and Validating Metamodels in Simulation", European Journal of Operational Research, vol. 20, pp. 14-29, 2000

[13] N. Goutal, C. Gœury, R. Ata, S. Ricci, N. El Mocyad, M. Rochoux, H. Oubanas, I. Gejadze, P.-O. Malaterre, "Uncertainty Quantification for River Flow Simulation Applied to a Real Test Case: the Garonne Valley", SimHydro 2017: Choosing the right model in applied hydraulics, Sophia Antipolis, 14-16 June 2017

[14] N. El Mocyad, S. Ricci, N. Goutal, M. Rochoux, S. Boyaval, C. Gœury, D. Lucor, O. Thual, "Polynomial Surrogates for Open-Channel Flows in Random Steady State", Environmental Modeling & Assessment, vol. 23(3), pp. 309–331, June 2018

[15] P. Roy, N. El Mocyad, S. Ricci, J.-C. Jouhaud, N. Goutal, M. De Lozzo, M. Rochoux, "Comparison of Polynomial Chaos and Gaussian Process surrogates for uncertainty quantification and correlation estimation of spatially distributed open-channel steady flows", Stochastic Environmental Research and Risk Assessment, vol. 32(6), pp. 1723-1741, April 2017

[16] S. El Garroussi, M. De Lozzo, S. Ricci, D. Lucor, N. Goutal, C. Gœury, S. Boyaval, "Uncertainty Quantification in a Two-Dimensional River Hyraulic Model", UNCECOMP 2019, 3[rd] ECOCOMAS Thematic Conference on Uncertainty Quantification in Computational Sciences and Engineering, Crete, Greece, 24-26 June 2019

[17] F. A. C. Viana, "Things you wanted to know about the Latin Hypercube Design and were afraid to ask", 10[th] World Congress on Structural and Multidisciplinary Optimization, Orlando, Florida, USA, pp. 1-9, 2013

[18] D. G. Krige, "A statistical approach to some mine valuation problems on the Witwatersrand", Journal of the Chemical, Metallurgical and Mining Society of South Africa, vol. 52, pp. 119-139, 1951

[19] G. Matheron, "Principles of Geostatistics", Economic Geology, vol 58(8), pp. 1246-1266, 1963

[20] I. M. Sobol, "Sensitivity estimates for nonlinear mathematical models", Mathematical Modelling and Computational Experiment, vol. 1, pp. 407–414, 1993